\documentclass[aps,pra,twocolumn,superscriptaddress]{revtex4}
\usepackage{graphicx}
\usepackage{subfig}
\usepackage{amssymb}
\usepackage{amsmath}
\usepackage{amsfonts}
\usepackage{bm}
\usepackage{color}
\usepackage{multirow}
\newcommand{\ket}[1]{\ensuremath{\left|{#1}\right\rangle}}

\newcommand{\oper}[1]{\mathbf{\mathsf{#1}}}

\begin{document}


\title{Detecting Entanglement of Continuous Variables with Three Mutually Unbiased Bases}

\author{E. C. Paul}
\affiliation{Instituto de F\'{\i}sica, Universidade Federal do Rio de
Janeiro, Caixa Postal 68528, Rio de Janeiro, RJ 21941-972, Brazil}
\author{D. S. Tasca}
\affiliation{Instituto de F\'{\i}sica, Universidade Federal do Rio de
Janeiro, Caixa Postal 68528, Rio de Janeiro, RJ 21941-972, Brazil}
\author{\L ukasz Rudnicki}
\affiliation{Institute for Theoretical Physics, University of Cologne,
Z{\"u}lpicher Stra{\ss}e 77 D-50937, Cologne, Germany}
\affiliation{Center for Theoretical Physics, Polish Academy of Sciences, Aleja
Lotnik\'ow 32/46, PL-02-668 Warsaw, Poland}
\author{S. P. Walborn}
\affiliation{Instituto de F\'{\i}sica, Universidade Federal do Rio de
Janeiro, Caixa Postal 68528, Rio de Janeiro, RJ 21941-972, Brazil}
\email[]{swalborn@if.ufrj.br}
\begin{abstract}
An uncertainty relation is introduced for a symmetric arrangement of three mutually unbiased bases in continuous variable phase space, and then used to derive a bipartite entanglement criterion based on the variance of global operators composed of these three phase space variables.  We test this criterion using spatial variables of photon pairs, and show that the entangled photons are correlated in three pairs of bases.  
\end{abstract}



\maketitle
\section{Introduction}
Entanglement detection in bipartite continuous-variable (CV) systems has typically been achieved using one out of several entanglement criteria involving measurement of two canonically conjugated variables, such as position and momentum. These criteria  are usually cast in terms of the variance \cite{duan00,mancini02}, covariance matrix \cite{simon00} or entropy \cite{walborn09,saboia11} of global variables, defined as linear combinations of the local variables.   
\par
In principle, bipartite separable quantum states can be perfectly correlated in one basis.  However, this perfect correlation implies that there is no correlation in the conjugate basis. Entangled quantum states, on the other hand, can be well correlated in both global position and global momentum variables.  The seminal example is the Einstein-Podolsky-Rosen (EPR) state, which as originally proposed is a simultaneous eigenstate of the relative position ($\oper{x}_1-\oper{x}_2$) and total momentum ($\oper{p}_1+\oper{p}_2$)  \cite{epr35}.     
\par
Of course it is possible to investigate quantum correlations in variables other than position and momentum.    For example, using dimensionless $x$ and $p$, one can consider rotated (dimensionless) operators of the form 
\begin{equation}
\oper{q}_\theta=\cos \theta \oper{x} + \sin \theta \oper{p}.
\label{eq:q}
\end{equation}

Using superscripts 1 and 2 to refer to the two parts of the bipartite system, it is well known that the EPR state is an eigenstate of the relative coordinate $\oper{q}^{1}_{\theta_1}-\oper{q}^{2}_{\theta_2}$ when $\theta_1+\theta_2$ is an integer multiple of $2 \pi$, and an eigenstate of the sum coordinate $\oper{q}^{1}_{\theta_1}+\oper{q}^{2}_{\theta_2}$ when $\theta_1+\theta_2$ is an odd multiple of $\pi$. In the context of transverse spatial correlations of photon pairs, this type of correlation was recently demonstrated experimentally in Refs. \cite{tasca08,tasca09a}, where the variances of canonically conjugate  $\oper{q}^{j}_{\theta_j}$ and $\oper{q}^{j}_{\theta_j^\prime}$ were chosen such that $\theta_j - \theta_j^\prime \equiv \pm \pi /2  \pmod{2\pi}$.  However, as will be discussed in the next section, it is not necessary to perform measurements satisfying this restriction.     

\par
The sets of eigenstates corresponding to canonically conjugate variables are examples of  \emph{mutually unbiased bases} (MUBs) \cite{durt10}. This means that, if a quantum system is an eigenstate of the observable--say--$\oper{x}$, then the probability distribution of the state with respect to the conjugate observable $\oper{p}$ is uniform. In other words, the precise knowledge of the state in one basis corresponds to complete ignorance in the conjugate basis.  In fact, as we discuss in more detail below, any two CV operators $\oper{q}_{\theta}$ and $\oper{q}_{\theta^\prime}$ of the form given in Eq. \eqref{eq:q}, with $\theta^\prime  \neq \theta \mod \pi$ define a pair of MUBs.      
Though it may have been known for some time that any pair of bases composed of the eigenstates of non-parallel (or non-antiparallel) phase space operators are MUBs,    
to our knowledge Weigert and Wilkinson \cite{weigert08} were the first to show that one can define up to three MUBs in the phase space of one CV bosonic mode.   
\par
In this work we investigate the use of a symmetric MUB triple for the investigation of quantum entanglement. We derive the relevant uncertainty relations for these variables and simple entanglement criteria based on the positive partial transpose (PPT) criterion \cite{peres96,horodecki96,simon00}, and use it to identify entanglement experimentally between two spatially entangled photons produced by Spontaneous Parametric Down Conversion (SPDC).     

\section{Mutually unbiased Phase Space Triple}
\label{sec:MUBTriple}
\begin{figure}
  \begin{center}
 \includegraphics{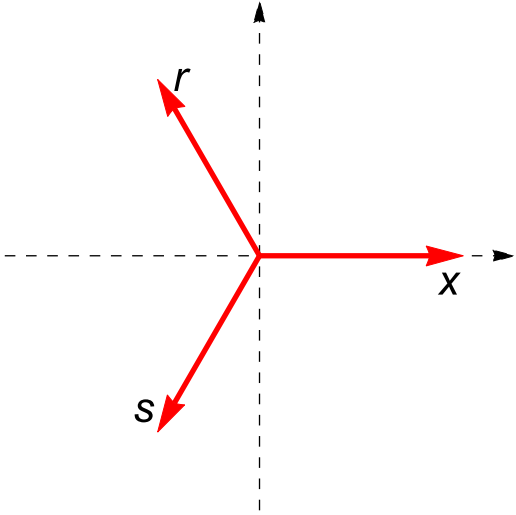}
  \caption{Three phase space variables $x$, $r$ and $s$ defining a mutually unbiased triple. Each variable is rotated $120^\circ$ from the other two.}
\label{fig:mub}
  \end{center}
  \end{figure}
Two bases $\{\ket{e} \}$ and $\{\ket{f}\}$ are mutually unbiased if all of their eigenstates have the same overlap $|\langle e| f \rangle |$. Position and momentum eigenstates are related via a Fourier transform, and consequently they satisfy $|\langle x| p \rangle |=1/ {\sqrt{2 \pi}}$ (we set $\hbar =1$). The position and momentum operators $\oper{x}$ and $\oper{p}$ satisfy the well-known Heisenberg-Robertson uncertainty relations \cite{rigolin15}
   \begin{equation}
\frac{1}{2}[  (\Delta \oper{x})^2 +  (\Delta \oper{p})^2] \geq  \Delta \oper{x} \Delta \oper{p} \geq  \frac{1}{2}|[\oper{x},\oper{p}]|=  \frac{1}{2}. 
\label{eq:URxp}
 \end{equation}
\par
Consider now two operators $\oper{q}$ and $\oper{q}^\prime$ given by Eq. \eqref{eq:q}.  These operators are related via a rotation in phase space, which is equivalent to a fractional Fourier transform (FRFT) \cite{ozaktas01,tasca11}.  Let us define $\theta_d=\theta^\prime-\theta$. Then we have
\begin{equation}
\oper{q}_{\theta^\prime}= \oper{F}^\dagger_{\theta_d}  \oper{q}_{\theta} \oper{F}_{\theta_d},
\end{equation} 
where $\oper{F}_{\theta_d}$ is the FRFT operator. Moreover, it is straightforward to show that 
    \begin{equation}
  \Delta \oper{q}_{\theta^\prime} \Delta \oper{q}_{\theta} \geq  \frac{1}{2}|[\oper{q}_{\theta^\prime},\oper{q}_{\theta}]|=  \frac{1}{2}|i \sin \theta_d|. 
\label{eq:URqqprime}
 \end{equation}
 
The scalar product between eigenstates gives the kernel to the FRFT \cite{ozaktas01},
 \begin{equation}
 \langle q_{\theta^\prime} | q_{\theta} \rangle = \sqrt{\frac{i e^{i\theta_d}}{2 \pi |\sin \theta_d|}} \exp \left [ i \frac{\cot \theta_d}{2} (q_{\theta}^2 + q_{\theta^\prime}^2) - i \frac{q_{\theta} q_{\theta^\prime}}{\sin \theta_d} \right ].
 \label{eq:FRFTkernel}
 \end{equation}
 
 One sees imediately that $|\langle q_{\theta^\prime} | q_{\theta}\rangle | = (2 \pi |\sin (\theta^\prime - \theta)|)^{-1/2}$, which does not depend on $q$ nor $q^\prime$, indicating that these two bases are mutually unbiased when $\sin(\theta^\prime-\theta) \neq 0$ \footnote{We note that proper limits of Eq. \eqref{eq:FRFTkernel} must be taken for the case $\sin (\theta^\prime - \theta) = 0$, returning  $\langle q^{\prime\prime}_{\theta} | q_{\theta} \rangle = \delta(q^{\prime\prime}_{\theta} - q_{\theta})$.}.
 
Recently, Weigert and Wilkinson \cite{weigert08} have shown that one can define three mutually unbiased bases in the phase space of one bosonic mode.  For example, consider the dimensionless operators $\oper{x}$, $\oper{r}$, and $\oper{s}$ corresponding to the phase space variables illustrated in Fig. \ref{fig:mub}. Explicitly, 
 \begin{equation}
 \oper{r} = \cos \frac{2 \pi}{3} \oper{x} + \sin \frac{2 \pi}{3} \oper{p},
 \label{eq:r}
 \end{equation}
 and 
  \begin{equation}
 \oper{s} = \cos \frac{4 \pi}{3} \oper{x} + \sin \frac{4 \pi}{3} \oper{p}.
  \label{eq:s}
 \end{equation}
 
These operators define a set of MUBs, since their eigenstates satisfy
 \begin{equation}
| \langle x \ket{r}\!| = | \langle x \ket{s}\!| = | \langle r \ket{s}\!| = \frac{1}{\sqrt{ 2 \pi \sin  \frac{2 \pi}{3}}}   = \frac{1}{\sqrt{\pi \sqrt{3}}}.
 \end{equation}
 
 
 Using Eqs. \eqref{eq:r} and \eqref{eq:s} and \eqref{eq:URqqprime}, these operators satisfy the uncertainty relations
 \begin{subequations}
 \label{eq:pairURs}
 \begin{equation}
\Delta \oper{x} \Delta \oper{r}  \geq \frac{1}{2}  \left|[\oper{x},\oper{r}]\right|  = \frac{1}{2} \left  |i \sin \frac{2 \pi}{3}\right | = \frac{\sqrt{3}}{4},
\end{equation}
\begin{equation}
\Delta \oper{x} \Delta \oper{s}  \geq \frac{1}{2} \left| [\oper{x},\oper{s}] \right| = \frac{1}{2} \left | i \sin \frac{4 \pi}{3}\right | = \frac{\sqrt{3}}{4},
\end{equation}
and 
\begin{equation}
\Delta \oper{r} \Delta \oper{s}  \geq \frac{1}{2} |[\oper{r},\oper{s}]| = \frac{1}{2}\left | i \sin \frac{2 \pi}{3} \right| =  \frac{\sqrt{3}}{4}.
\end{equation}
\end{subequations}


We note that, up to a rotation of the entire phase-space, $x$, $r$ and $s$ constitute the only possible normalized set of three MUBs. Nevertheless, if we allow for individual scaling of the variables, there are other sets of three variables that satisfy the condition of MUBs \cite{weigert08}. Here we focus on the operators defined in Eqs. \eqref{eq:r} and \eqref{eq:s}, since they can be obtained by simple rotations in phase space, which can be achieved experimentally with relative ease in a number of systems \cite{tasca08,tasca09a,barbosa13a,barbosa13b}. 
\subsection{Uncertainty relations for the symmetric phase space triple}
\label{sec:3UR}
A number of uncertainty relations can be derived for the variables shown in Fig. \ref{fig:mub} a).   Some of these follow trivially from the usual uncertainty relations for pairs of operators \eqref{eq:pairURs}.  For example,  taking the product of all three pairwise URs \eqref{eq:pairURs} gives 
\begin{equation}
(\Delta \oper{x})^2  (\Delta \oper{r})^2  (\Delta \oper{s})^2   \geq \frac{3 \sqrt{3}}{64} \approx 0.08, 
\label{eq:VTPbad}
\end{equation}
which does not appear to be tight, since for the vacuum state the triple product above is $1/8=0.125$.    
\par
We will now show how one can arrive at a tight lower bound for the triple product $(\Delta \oper{x})^2 (\Delta \oper{r})^2 (\Delta \oper{s})^2 $.  First, let us use Eqs. \eqref{eq:r} and \eqref{eq:s} and the definition of variance to write
\begin{equation}
(\Delta \oper{r})^2  = \frac{1}{4}(\Delta \oper{x})^2  +  \frac{3}{4}(\Delta \oper{p})^2  - \frac{\sqrt{3}}{4}\langle\{ \oper{x}, \oper{p}\} \rangle+ \frac{\sqrt{3}}{2}\langle \oper{x}\rangle \langle \oper{p}\rangle,
\label{eq:Varr}
\end{equation}
and
\begin{equation}
(\Delta \oper{s})^2  = \frac{1}{4}(\Delta \oper{x})^2  +  \frac{3}{4}(\Delta \oper{p})^2  + \frac{\sqrt{3}}{4}\langle\{ \oper{x}, \oper{p}\}\rangle - \frac{\sqrt{3}}{2}\langle \oper{x}\rangle \langle \oper{p}\rangle ,
\label{eq:Vars}
\end{equation}
which multiplied together give
\begin{align}
(\Delta \oper{r})^2  (\Delta \oper{s})^2  = & \frac{1}{16}[(\Delta \oper{x})^2  +  3(\Delta \oper{p})^2 ]^2 - \nonumber \\ & \frac{3}{16}\left( \langle\{ \oper{x}, \oper{p}\}\rangle - 2 \langle \oper{x}\rangle \langle \oper{p}\rangle \right)^2.
\label{eq:VTP2}
\end{align}

The Schr\"odinger-Robertson UR for operators $\oper{x}$ and $\oper{p}$, which is a state-dependant generalization of \eqref{eq:URxp}, reads \cite{rigolin15}
\begin{equation}
(\Delta \oper{x})^2 (\Delta \oper{p})^2  \geq \frac{1}{4} + \frac{1}{4}\left( \langle\{\oper{x}, \oper{p}\}\rangle - 2 \langle \oper{x}\rangle \langle \oper{p}\rangle \right)^2.  
\end{equation}
which can be rewritten in the form
\begin{equation}
- \left( \langle\{\oper{x}, \oper{p}\}\rangle  - 2 \langle \oper{x}\rangle \langle \oper{p}\rangle \right)^2 \geq 1 - 4 (\Delta \oper{x})^2 (\Delta \oper{p})^2 .  
\end{equation}

Using this expression in Eq. \eqref{eq:VTP2}, and performing some algebra, we have
\begin{equation}
(\Delta \oper{r})^2  (\Delta \oper{s})^2  \geq \frac{3+((\Delta \oper{x})^2  -  3(\Delta \oper{p})^2 )^2}{16},
\label{eq:VTP4new}
\end{equation} 
which means the triple product of the variances is then
\begin{equation}
(\Delta \oper{x})^2  (\Delta \oper{r})^2  (\Delta \oper{s})^2  \geq (\Delta \oper{x})^2  \frac{3+[(\Delta \oper{x})^2  -  3(\Delta \oper{p})^2 ]^2}{16}.  
\label{eq:VTP4}
\end{equation}

Optimizing the right-hand side over all positive values of the $x$ and $p$ variances respecting the Heisenberg UR $(\Delta \oper{x})^2  (\Delta \oper{p})^2  \geq 1/4$ gives (see Appendix A)
\begin{equation}
(\Delta \oper{x})^2  (\Delta \oper{r})^2  (\Delta \oper{s})^2  \geq \frac{1}{8}.
\label{eq:VTP}
\end{equation}

Contrary to the usual Heisenberg uncertainty relation, which can also be saturated by Gaussian squeezed states, inequality \eqref{eq:VTP} is saturated exclusively by the set of coherent states, as we show in Appendix A.        

  \section{Entanglement Criterion with Triples}
  \label{sec:EntCrit}
  Quantum mechanical uncertainty relations combined with the positive partial transpose argument \cite{peres96,horodecki96} can be used to derive entanglement criteria \cite{simon00,walborn09,toscano15}.  
  Let us define the global phase space operators
  \begin{subequations}
  \label{eq:globalops}
  \begin{align}
  \oper{X}_{\pm} & = \oper{x}_1 \pm \oper{x}_2, \\
   \oper{R}_{\pm} & = \oper{r}_1 \pm \oper{r}_2, \\
   \intertext{and}
   \oper{S}_{\pm} & = \oper{s}_1 \pm \oper{s}_2.
  \end{align}
  \end{subequations}
  
  Note that we can also write 
  \begin{equation}
  \oper{R}_{\pm} = \cos \frac{2 \pi}{3} \oper{X}_{\pm}  + \sin \frac{2 \pi}{3} \oper{P}_{\pm},
  \label{eq:R}
  \end{equation}
  and 
  \begin{equation}
  \oper{S}_{\pm} = \cos \frac{4 \pi}{3} \oper{X}_{\pm} +  \sin \frac{4 \pi}{3} \oper{P}_{\pm},
  \label{eq:S}
  \end{equation}
  where $\oper{P}_{\pm} = \oper{p}_{1} \pm \oper{p}_2$, and we note that $[\oper{X}_{\pm},\oper{P}_{\pm}] = 2i$ and $[\oper{X}_{\pm},\oper{P}_{\mp}] =0$.  Following the derivations in section \ref{sec:3UR}, it is straightforward to show that the global operators \eqref{eq:globalops} satisfy the uncertainty relation
\begin{equation}
(\Delta \oper{X}_\pm)^2  (\Delta \oper{R}_\pm)^2  (\Delta \oper{S}_\pm)^2  \geq 1, 
\label{eq:URGlobalVTP}
\end{equation}
where either the top row of all plus signs or the bottom row of all minus signs is considered. These URs can be used to develop entanglement criteria following arguments previously given in Refs. \cite{agarwal05,walborn09,saboia11,tasca13,toscano15}.
 
Naturally, any bipartite state $\rho_{12}$ satisfies either of the uncertainty relations  \eqref{eq:URGlobalVTP}. Now suppose its partial transpose with respect to--say--system two, $\rho_{12}^{T_2}$, is positive. In this case we can say for sure that $\rho_{12}^{T_2}$ is a bonafide quantum state and also satisfies the uncertainty relations \eqref{eq:URGlobalVTP}. Following this argument, the PPT criterion establishes that if a bipartite state $\rho_{12}$ has a negative partial transpose, then it cannot be separable and must be entangled \cite{peres96,horodecki96}. For continuous variables, Simon has shown that the partial transpose is equivalent to a mirror reflection, transforming the  phase space variable $p_2 \longrightarrow -p_2$ and leaving all others unchanged \cite{simon00}.  This means taking the global variable $P_{\mp} \longrightarrow P_{\pm}$.

Let us now define two new global operators
 \begin{equation}
 \oper{U}_{\pm}  = \cos \frac{2 \pi}{3} \oper{X}_{\pm}  + \sin \frac{2 \pi}{3} \oper{P}_{\mp}
 = \oper{r}_{1} \pm \oper{s}_{2}
  \label{eq:RT} 
  \end{equation}  
and 
  \begin{equation}
  \oper{V}_{\pm} = \cos \frac{4 \pi}{3} \oper{X}_{\pm} +  \sin \frac{4 \pi}{3} \oper{P}_{\mp} = \oper{s}_{1} \pm \oper{r}_{2}.
  \label{eq:ST}
 \end{equation}
 
The standard deviations (square roots of variances) of operators $\oper{X}_{\pm}$, $\oper{U}_{\pm}$ and $\oper{V}_{\pm}$ for the original state can be related to the standard deviations  of variables $\oper{X}_{\pm}$, $\oper{R}_{\pm}$ and $\oper{S}_{\pm}$  on the transposed state as $ \Delta \oper{X}_{\pm,T}=\Delta \oper{X}_\pm $, $ \Delta \oper{R}_{\pm,T}=\Delta \oper{U}_\pm$, $\Delta \oper{S}_{\pm,T}= \Delta \oper{V}_\pm$, where again $T$ stands for partial transposition. Thus, using the UR  \eqref{eq:URGlobalVTP}, any separable state $\rho_{12}$ will satisfy the inequalities
      \begin{equation}
(\Delta \oper{X}_\pm)^2  (\Delta \oper{U}_\pm)^2  (\Delta \oper{V}_\pm)^2  \geq 1.
\label{eq:EntCrit2}
\end{equation}

 A state that violates either of the inequalities  \eqref{eq:EntCrit2} has a negative partial transpose and is thus entangled. Moreover, the operators $ \oper{X}_{j}$, $\oper{U}_{j}$ and $\oper{V}_{j}$ ($j=+$ or $ -$) commute.  This means they share a common eigenstate, namely, the EPR state \cite{epr35}, for which all of the variances are zero. 
 
Let us now give a real world example of a quantum state that violates criteria \eqref{eq:EntCrit2}.  Under appropriate conditions, the spatial variables of photon pairs produced from SPDC are well described by the double Gaussian wave function \cite{law04,walborn10,schneeloch15}
\begin{equation}
\Psi(x_1,x_2) =  A \exp \left (-\frac{(x_1+x_2)^2}{4 \sigma_+^2} \right) \exp \left (-\frac{(x_1-x_2)^2}{4 \sigma_-^2} \right),
\label{eq:state}
\end{equation}
where $A=1/\sqrt{\pi \sigma_+\sigma_-}$ and the variables $x$ refer to the transverse position variables at the exit face of the nonlinear crystal.  Here we consider the simple case of one spatial dimension.  This state is entangled when $\sigma_- \neq \sigma_+$.  The SPDC state \eqref{eq:state} is analogous to the two-mode squeezed state when $\sigma_- = 1/\sigma_+$.  Though the EPR state mentioned above is unphysical, it is a limiting case of the two-mode squeezed state in the case of infinite squeezing, when $\sigma_- \longrightarrow 0$. 
\par
Under usual experimental conditions, we have $\sigma_- << \sigma_+$ and the two-photon state shows position correlation at the exit face of the crystal.  Using lenses or free-propagation, it is possible to observe correlations in other phase space variables.  For example, an optical Fourier transform system allows one to observe the momentum anti-correlations of this state \cite{howell04}.  Using other optical systems, correlations in other (rotated) variables can be observed.  It has been shown \cite{tasca08,tasca09a} that entangled states of the form \eqref{eq:state} with $\sigma_- < \sigma_+$ display position correlations when the sum of the phase-space rotation angles $\theta_1$ and $\theta_2$, with respect to $x_{1}$ and $x_{2}$, respectively, is an even multiple of $\pi$, and anti-correlation when the sum is an odd multiple of $\pi$. From the definitions \eqref{eq:r} and \eqref{eq:s}, the $r$ variables are given by a rotation of $2\pi/3$ and the $s$ variables by a rotation of $4 \pi/3$. Noting that $\oper{U}_{\pm} = \oper{r}_1\pm \oper{s}_2$ and   $\oper{V}_{\pm} = \oper{s}_1\pm \oper{r}_2$, we can see that $\theta_{1} + \theta_{2} = 2 \pi$ in both cases. Since $x_{1}$ and $x_{2}$ are also correlated, with $\theta_{1} + \theta_{2} = 0$, the state \eqref{eq:state} should show correlations in all three sets of variables, leading to a violation of the entanglement criteria \eqref{eq:EntCrit2}.     
 \section{Experiment and Results}
 
 \begin{figure}
  \begin{center}
\includegraphics[width=8cm]{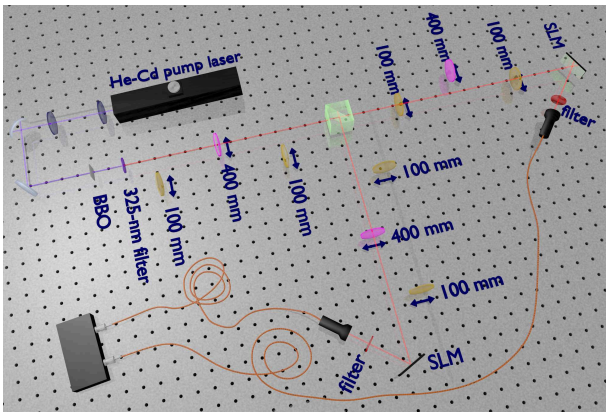}
  \caption{Experimental setup. The SLMs are spatial light modulators used to automatically scan a slit across the transverse distribution of the down-converted photons.}
\label{fig:setup}
\end{center}
  \end{figure}
  
To test the entanglement criteria derived, we observed the transverse spatial variables of the twin-photon state generated by the process of SPDC \cite{walborn10}.  To generate the state, a 2-mm thick beta-barium borate (BBO) crystal cut for type-I phase matching SPDC was pumped by a continuous-wave 325-nm  He-Cd laser beam, producing collinear degenerate converted beams at a wavelength of 650 nm.  The two down-converted beams were separated at a 50/50 beam splitter and each directed by mirrors and lenses to Holoeye Pluto phase-only spatial light modulators (SLM), which were used to perform position correlation measurements by scanning a phase slit in the transverse profile of the beams, as described in Ref. \cite{davis99}.

Both beams were scanned in the horizontal direction of the transverse detection planes over a region of interest of 12 mm. Using slits of 80 $\mu$m (equivalent to 10 pixels of our SLM), this procedure totalled $150 \times 150 = 22 500$ data points per measurement. Each data point was sampled for 3 s, leading to the estimated joint detection probabilities.
 
 After reflection by the SLMs, the beams were sent through 10-nm FWHM interference filters centered at 650 nm, and were then coupled into multi-mode optical fibers connected to single-photon avalanche diodes (SPAD).

Non-confocal lenses were used in an optical fractional Fourier transform arrangement \cite{tasca08,tasca09a,ozaktas93,lohmann93,ozaktas01} to achieve the phase-space rotations described in section \ref{sec:MUBTriple}. As can be seen in the sketch of our experimental setup displayed in Fig. \ref{fig:setup}, between the BBO crystal and the beam splitter either two 100-mm confocal lenses were used to achieve a phase-space rotation of $\pi$, or a single 400-mm lens was used to perform a $\pi/3$ rotation. After the beam splitter, similar sets of lenses were used in order to achieve $2\pi$, $2\pi/3$ or $4\pi/3$ in each converted beam, depending on the lenses chosen. 
In order to interpret the fractional Fourier transform as a rotation, dimensionless variables must be used. Moreover, it is important that the focal lengths of each lens system are chosen so that the scaling factor is the same for each transformation.  The measured dimensional position variables on the SLM plane were converted to dimensionless variables through the scaling factor $d=\sqrt{f \sin(\pi/3)/k}$, where $f$ is the focal length of the lens, $\theta$ is the phase-space rotation angle and $k$ is the wavenumber of the down-converted beams \cite{tasca11}. In this experiment, lens systems were chosen so as to allow for the same scaling factor $d=189 \mu$m for the three measurements.
  \begin{figure}
    \centering
    \subfloat{{\includegraphics[width=4cm]{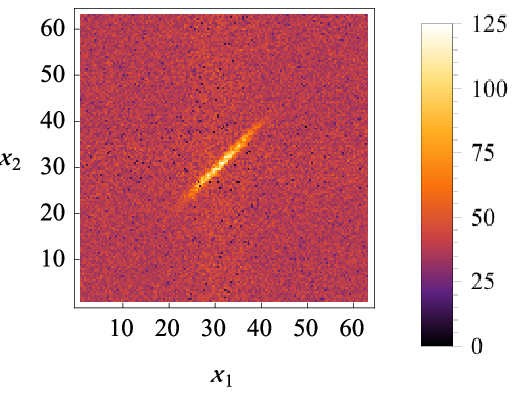} }}
    \quad
    \subfloat{{\includegraphics[width=4cm]{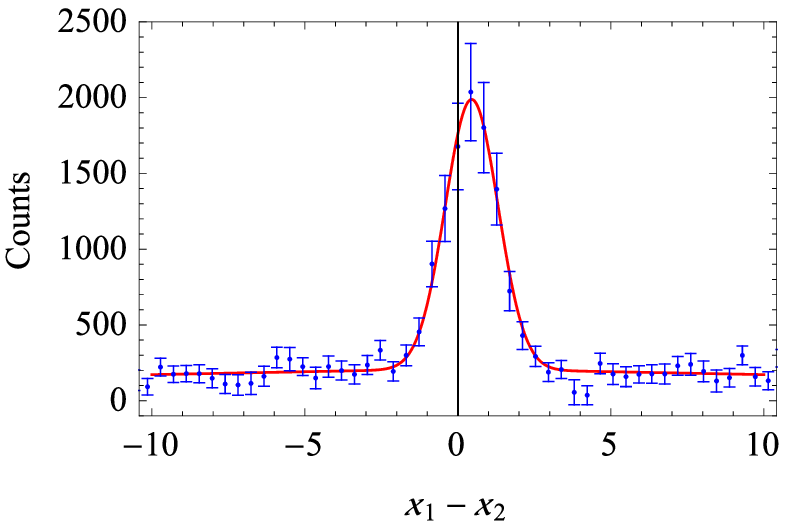} }}
    \quad \\
   \subfloat{{\includegraphics[width=4cm]{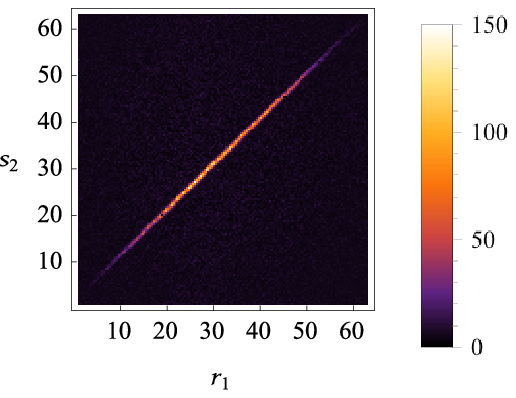} }}
    \quad
    \subfloat{{\includegraphics[width=4cm]{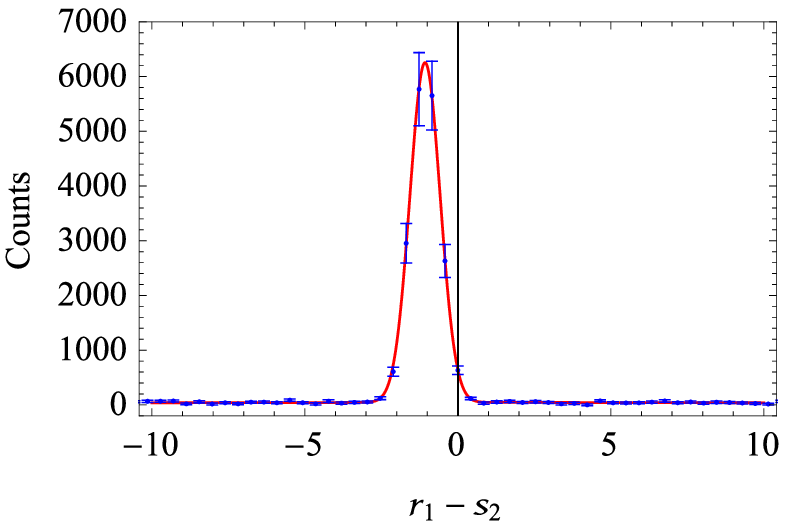} }}
    \quad \\
    \subfloat{{\includegraphics[width=4cm]{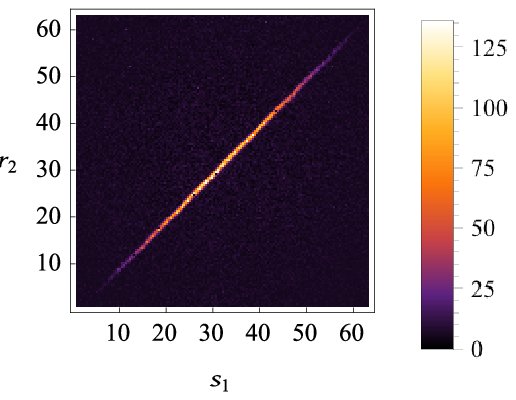} }}
    \quad
    \subfloat{{\includegraphics[width=4cm]{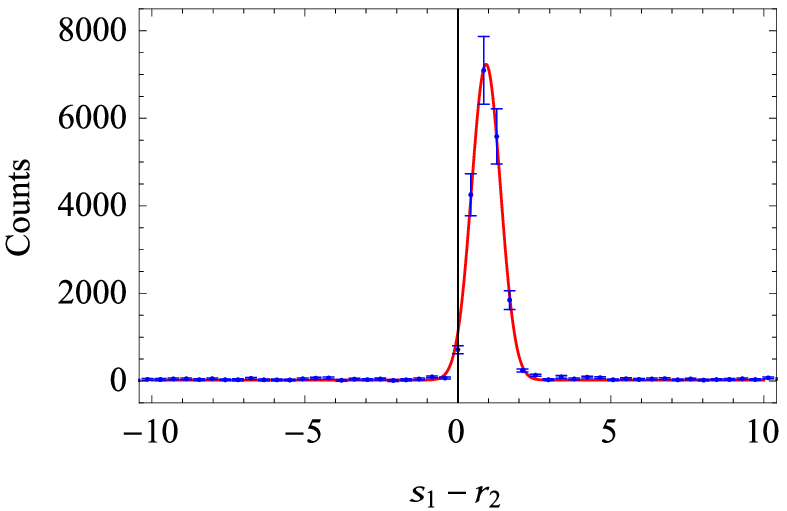} }}
    \caption{Measurement results. On the left-hand side we can see the reconstructed joint distributions of the dimensionless variables as obtained experimentally, and on the right-hand side are their marginal distributions. The curves shown correspond to Gaussian best fits. The error bars correspond to the square root of the photon counts, due to the fact that the photon-count distribution is poissonian.}
    \label{fig:results}
\end{figure}

The measurement results can be seen in Fig. \ref{fig:results}. On the left-hand side the reconstructed joint distributions of the dimensionless variables are shown, from which one can clearly see the intensity correlations in all three pairs of spatial variables.  We note that the usual momentum anti-correlations are never observed, though they play a crucial role in the $r_1,s_2$ and $s_1,r_2$ correlations, as can be seen in Eqs. \eqref{eq:RT}  and \eqref{eq:ST}. 
The blue points in the plots on the right-hand side are the marginal distributions of the relevant global variables.  The error bars correspond to error due to Poissonian count statistics, which attributes a standard deviation equal to the square root of the count rate. 
The red curves correspond to Gaussian best fits of the marginal distributions, from which we are able to obtain the variances, which are shown in the second column of of Table \ref{tab:1}.   
\begin{table}
\begin{tabular}{|c|c|c|c|}
\hline
$W$ & $(\Delta \oper{W}_{-})^2$ & $(\Delta \oper{W}_{+})^2$ & $C_W$ \\ 
\hline
$X$ &  $0.74 \pm 0.02$ & $321 \pm 29$ & $21 \pm 2$ \\
\hline
$U$ &  $0.2455 \pm 0.0006$ & $554 \pm 69$ & $55 \pm 3$\\
\hline
$V$ & $0.225 \pm 0.001$ & $598 \pm 66$ & $52 \pm 3$ \\
\hline
\end{tabular}
\caption{\label{tab:1} Variances of global variables $W_{\pm}$ ($W=X,U,V$) obtained from gaussian fits of marginal distributions of coincidence plots shown in Fig. \ref{fig:results}.  The last column is the correlation coefficient, defined in the text.}
\end{table}
With these results, inequality  \eqref{eq:EntCrit2} is
\begin{equation}
(\Delta \oper{X}_{-})^2 (\Delta \oper{U}_{-})^2 (\Delta \oper{V}_{-})^2 
 = (0.041 \pm 0.001) \ngeq 1,
 \end{equation}
showing that the correlations in the state are sufficient to violate the inequality. Obviously the variance $(\Delta \oper{X}_{-})^2$ is larger than the other two. This is to be expected, as one can see from inspection of Eqs. \eqref{eq:RT} and \eqref{eq:ST} that the variances in the rotated global variables in fact depend on both the variance in the near-field as well as the far-field.  
Indeed, since our state is not symmetric, meaning $\sigma_- \neq 1/\sigma_+$ in the wavefunction \eqref{eq:state},  we do not expect $(\Delta \oper{X}_{-})^2$ to be equal to $(\Delta \oper{U}_{-})^2$ nor  $(\Delta \oper{V}_{-})^2$.   
 However, we should see the same amount of correlation in each set of measurements, as we will now explain.  Following \cite{schneeloch15}, let us define a correlation coefficent $C_{W}$ ($W=X,U,V$) as $C_W= {\Delta \oper{W}_{+}}/{\Delta \oper{W}_{-}}$.  Using the double Gaussian wavefunction in Eq. \eqref{eq:state}, together with the variances defined in Eqs. \eqref{eq:RT} and \eqref{eq:ST}, and performing a little algebra we find
\begin{equation}
C_U = C_V= \sqrt{ \frac{\sigma_+^2+3/\sigma_-^2}{\sigma_-^2+3/\sigma_+^2}}=\frac{\sigma_+}{\sigma_-}, 
\end{equation}
which is exactly the correlation ratio $C_X= {\Delta \oper{X}_{+}}/{\Delta \oper{X}_{-}}$.   The value $\sigma_+/\sigma_-$ is related to the Schmidt coefficient of the two-photon state, and thus to the amount of entanglement.  The values of the variances in the sum coordinates $(\Delta \oper{W}_{+})^2$, as well as the correlation coefficients are shown in Table \ref{tab:1}.  One can see from the table  that we see less correlation in the near-field ($x_1,x_2$) measurements, by roughly a factor of 2.5 compared to the measurements in the other planes.  This is also observable in the coincidence maps of Fig. \ref{fig:results}.  The reduced correlation in the near-field measurements is in most part due to extra noise that appears (see Fig. \ref{fig:results}) owing to fluorescence of the laser beam on the dichroic filter, which is then imaged onto the detection planes. The rotated variables do not suffer from this noise as drastically, which is an advantage to not working in the image plane of the source \cite{tasca08}.

 \section{Conclusion}
 We have derived uncertainy relations and entanglement criteria for continuous variables using three mutually unbiased bases. By measuring the spatial correlations between photon pairs generated by SPDC we were able to show that the photons were correlated in three pairs of variables, and that these correlations lead to violation of a separabilty criterion. Considering entanglement detection in three mutually unbiased bases could be interesting for quantum key distribution, and might improve the sensitivity to an eavesdropper.  

 \acknowledgements
 
ECP, DST and SPW acknowledge financial support from the Brazilian funding agencies CNPq, FAPERJ, CAPES, and the National Institute for Science and Technology - Quantum Information. DST acknowledges FAPERJ for financial support under grant No.  E-26/101.234/2013 and E-26/101.264/2013.
 {\L}.R. acknowledges financial support from Grant No. 2014/13/D/ST2/01886 of
the Polish National Science Centre. Research in Cologne
is supported by the Excellence Initiative of the German Federal and State Governments (Grants ZUK 81), the ARO under contracts W911NF-14-1-0098 and W911NF-14-1-0133 (Quantum Characterization, Verification, and Validation), and the DFG (GRO 4334/2-1). {\L}.R. acknowledges hospitality of Freiburg Center for Data Analysis and Modeling.

\appendix
\section{Derivation of Eq. \ref{eq:VTP}}
Let us introduce the shorthand notation: $\eta=\left(\Delta x\right)^{2}$ and $\xi=\left(\Delta p\right)^{2}$.
According to (\ref{eq:VTP4}), we aim to find a minimum of the function
\begin{equation}
g\left(\eta,\xi\right)=\frac{\eta}{16}\left[3+(\eta-3\xi)^{2}\right],
\end{equation}
given the inequality constraint $\eta\xi\geq1/4$ coming from the Heisenberg Uncertainty Relation (HUR). 

Consider first the sharp-inequality case $\eta\xi>1/4$. We can calculate the derivatives
\begin{equation}
\frac{\partial g}{\partial\eta}=\frac{3}{16}\left(1+\eta^{2}-4\eta\xi+3\xi^{2}\right),
\end{equation}
\begin{equation}
\frac{\partial g}{\partial\xi}=\frac{3\eta}{8}\left(3\xi-\eta\right),
\end{equation}
and easily see that the system of equations
\begin{equation}
\frac{\partial g}{\partial\eta}=0,\qquad\frac{\partial g}{\partial\xi}=0,
\end{equation}
has no solutions.

In the second case, when the HUR is saturated, the function to be minimized becomes
\begin{equation}
g_{\textrm{sat}}\left(\eta\right)\equiv g\left(\eta,1/4\eta\right)=\frac{\eta}{16}\left[3+\left(\eta-\frac{3}{4\eta}\right)^{2}\right].
\end{equation}
We find
\begin{equation}
\frac{dg_{\textrm{sat}}}{d\eta}=\frac{3}{256}\left(16\eta^{2}-\frac{3}{\eta^{2}}+8\right),
\end{equation}
and this derivative is equal to $0$ when $\eta=\pm1/2$ or $\eta=\pm i\sqrt{3}/2$.
Since the parameter $\eta$ is real and non-negative, we are left
with the single solution $\eta=1/2$. This solution corresponds to
the Gaussian coherent state as in this case also $\xi=1/2$. Obviously
$g\left(1/2,1/2\right)=1/8$. Finally, we check that the second derivative
\begin{equation}
\frac{d^{2}g_{\textrm{sat}}}{d\eta^{2}}=\frac{9}{128\eta^{3}}+\frac{3\eta}{8},
\end{equation}
is always positive, so the solution is a true minimum.  
\par
One might imagine some non-gaussian state with appropriate symmetry, or gaussian squeezed state that saturates the triple product UR.  However, the solution above dictates explicitely that $(\Delta x)^2 =  (\Delta p)^2=1/2$, which is only attainable by the vacuum state and by displaced vacuum states--the set of coherent states.  In other words, there are no non-gaussian states that saturate the triple product UR.  Moreover, it is impossible to apply a squeezing operation such that the variances in both the $x$ and $p$ (or any perpendicular directions) remain equal to $1/2$.      

 \bibliographystyle{apsrev}

\end{document}